# SIMULATION OF THE FAST ION INSTABILITY IN SSRF STORAGE RING

Guoxing Xia[1], Bocheng Jiang[2], Guimin Liu[2]
1. Max Planck Institute for Physics, 80805, Munich, Germany
2. Shanghai Institute of Applied Physics, CAS, Shanghai, China

*Abstract*

Fast ion instability has been observed in the early commissioning and operation of the Shanghai Synchrotron Radiation Facility (SSRF) storage ring. In this paper, a weak-strong code is used to simulate the fast ion instability in SSRF storage ring. Various fill patterns and gas pressures are investigated. The results show that the mini-train fill patterns are very effective to suppress the growth of the fast ion instability.

## INTRODUCTION

In a storage ring, when the charged particle beam passes through an accelerator beam pipe, the ions produced via collision ionization, synchrotron radiation ionization and tunnel ionization in the residual gas in the beam pipe will interact with the particle beam. These ions have long been recognized as a potential limitation in the machines like electron and antiproton storage rings. They may result in beam emittance growth, tune shifts and beam lifetime reduction etc.

There are two kinds of ion effects. One is classical ion trapping instability in which the ions are accumulated over many turns in the ring and trapped by the beam potential. This instability can be cured by intentionally leaving a gap after the bunch train. The gap makes these ions over-focused and lost at the vacuum chamber wall. In addition, the growth time of ion trapping instability is long enough which can be suppressed by introducing a fast feedback system. Meanwhile reducing the gas pressure in the vacuum pipe also helps to alleviate the growth of the ion trapping instability.

However, in many modern storage rings, the extremely small beam emittance (nanometer scale) and a train of many bunches (few hundreds to thousands bunches) are featured to increase the machines' brilliance or luminosity, effect of single-passage ions called fast ion instability (FII) is therefore noticeable [1,2]. In this case, the ions are generated during a single passage of the bunch train and the ion density increases roughly linearly along the bunch train. The number of ions produced by a single long bunch train is large enough to disturb the beam motion at tail of the train. In the gap at the end of the bunch train the most of the ions are lost and left an insignificant number of ions when the bunch train returns on its next revolution. This phenomenon has already observed in many electron storage rings and ring-based light sources, e.g., ALS, PLS, TRISTAN-AR, ELETTRA, ATF damping ring etc [3].

The Shanghai Synchrotron Radiation Facility (SSRF) is an intermediate energy storage ring based light source [4]. It was designed as one of the advanced third generation light sources worldwide. The design emittance of the SSRF storage ring is about 3.9 nm·rad and the beam current is above 200 mA. The basic beam parameters of the SSRF storage ring are listed in Table 1. During the early machine commissioning and operation, the fast ion instability was observed, which shows the beam size growth has a strong dependence on the bunch index. To investigate this instability in more detail, in this paper, a weak-strong simulation code is employed to simulate the interactions between the ions and the electrons in SSRF storage ring.

Table 1: The beam parameters of the SSRF storage ring

| | |
|---|---|
| Energy (GeV) | 3.5 |
| Circumference (m) | 432 |
| Harmonic number | 720 |
| Natural emittance (nm) | 3.9 |
| Transverse coupling | 0.7% |
| Beam current (mA) | $\geq 200$ |
| Bunch spacing (RF bucket) | 1 |
| Betatron tunes ($Q_x/Q_y$) | 22.22/11.29 |
| Synchrotron tune $Q_s$ | $7.2\times10^{-3}$ |
| Momentum compaction | $4.27\times10^{-4}$ |
| Natural chromaticities ($\xi_x/\xi_y$) | -55.7/-17.9 |
| Relative energy spread | $9.7\times10^{-4}$ |
| RF frequency (MHz) | 499.654 |
| Damping times $\tau_x/\tau_y/\tau_z$ (ms) | 7.35/7.36/3.68 |

## SIMULATION ALGORITHM

Simulation study of the FII was begun by T. Raubenheimer and F. Zimmermann [1]. Here we used a more simplified simulation method for reducing the computation time. This simulation is based on a weak-strong model [5], which allows us to track the bunch motion in a long term period (more than one damping time).

In our simulation, the beam is assumed to be Gaussian distribution in the transverse plane. Only the barycentre motion of each bunch is taken into account. The produced ions are regarded as the macro-particles. The equations of motion for bunches and ions are expressed by

$$\frac{d^2 x_{e,a}}{ds^2} + K(s) x_{e,a} = \frac{2r_e}{\gamma} \sum_{j=1}^{N_i} F(x_{e,a} - x_{i,j}; \sigma(s))\delta(s-s_i) \quad (1)$$

$$\frac{d^2 x_{i,j}}{dt^2} = \frac{2N_e r_e c^2}{M_i/m_e} \sum_{a=1}^{N_b} F(x_{i,j} - x_{e,a}; \sigma(s))\delta(t-t(s_{e,a})) \quad (2)$$

here $x_{e,a}$ and $x_{i,j}$ are the dipole momenta for electron beam and ions, $\sigma$ is the beam size, $N_b$, $N_i$, $M_i$ and $m_e$ are the number of bunches, number of ions, mass of ion and electron respectively, $r_e$ and $c$ are the classical radius of electron and the velocity of light, $\gamma$ is the Lorentz factor of

the beam. $F(x;y)$ is the well known Bassetti-Erskine formula which is given by [6]

$$f(x;y) = -\frac{\sqrt{\pi}}{\sqrt{2(\sigma_x^2 - \sigma_y^2)}} \left[ w\left(\frac{x+iy}{\sqrt{2(\sigma_x^2 - \sigma_y^2)}}\right) - \exp\left(-\frac{x^2}{2\sigma_x^2} - \frac{y^2}{2\sigma_y^2}\right) w\left(\frac{\frac{x\sigma_y}{\sigma_x} + i\frac{y\sigma_x}{\sigma_y}}{\sqrt{2(\sigma_x^2 - \sigma_y^2)}}\right) \right]$$

in which $w(z) = \exp(-z^2)[1 - erf(-iz)]$ and the error function is given by

$$erf(x) = \frac{2}{\sqrt{\pi}} \int_0^x \exp(-x^2)dx$$

here $\sigma_x$ and $\sigma_y$ are horizontal and vertical beam sizes respectively. Simulations of two beam instability are performed by solving the differential Eqs. (1) and (2).

In the simulation, the number of ions is increased with respect to the number of bunches in the bunch train. The first bunch in the train only produces the ions and it does not interact with the ions. The following bunches interact with the ions produced by their preceding bunches. After one turn interaction, we assume that the ions are cleared away from the beam vicinity (by clearing gap or clearing electrodes). The new ions will be produced by the beam in the second revolution turn. In our simulation, the real lattice is considered in which the beta function and dispersion function vary in different locations along the ring. To save CPU time of simulation, 20 interaction points along the ring are used. In these interaction points, we artificially enhance the number of ions by taking into account the real vacuum pressure of the ring. The adjacent beam-ion interaction points are connected through the linear transfer matrix. The beam parameters used here are taken from Table 1.

## SIMULATION RESULTS

For the SSRF storage ring, the vertical beam emittance is much smaller than the horizontal one (coupling factor of 0.7%), the FII is much more significant in the vertical plane, as we can see from the following simulation results (Fig.1). In our simulations, the time evolution of the growth of beam dipole amplitude is simulated and recorded turn by turn. The data are recorded for 5000 turns which is around one damping time. The vertical oscillation amplitude of the bunch centroid is half of the Courant-Synder invariant and given by

$$J_y = [\gamma y^2 + 2\alpha y y' + \beta y'^2]/2$$

where α, β and γ are the Twiss parameters of the ring determined by the ring lattice. The value of $\sqrt{J_y}$ is compared with the vertical beam size which is represented by the value of $\sqrt{\varepsilon_y}$ (here $\varepsilon_y$ is the beam vertical emittance).

The major species of the residual gas in the vacuum chamber are Carbon Monoxide (CO) and Hydrogen ($H_2$). Since the cross section of collision ionization for CO is about 6 times higher than that for $H_2$ in this beam energy regime. Therefore in the simulation, $CO^+$ ions are regarded as the dominant instability source.

Fig. 1 shows beam maximum horizontal oscillation amplitude with respect to the number of turns for a typical fill pattern in which 600 bunches with bunch current of 0.5 mA are filled in the ring. Here we assume the CO partial pressure of 1.0 nTorr. In this figure, $N_0$ denotes the number of particles per bunch, $n_b$ the number of bunches per train and $n_{train}$ is the number of trains. It can be seen that the beam oscillation amplitude is well below the beam size in this case (horizontal beam size is around $6 \times 10^{-5}$ m). Fig.2 gives the beam maximum vertical oscillation amplitude with respect to the number of turns with the same parameters as used in Fig.1. It indicates that the beam oscillation amplitude increases with respect to the number of turns. For the CO pressure of 1.0 nTorr, the growth of vertical amplitude is beyond the beam size, depicted by a dot line in the figure. Fig. 3 gives the growth of maximum oscillation amplitude with respect to the number of turns for another typical fill pattern in which 450 bunches with bunch current of 0.3 mA are injected in the ring. It shows that the vertical amplitude of beam is also beyond the beam size after the $3830^{th}$ turn at CO pressure of 1.0 nTorr. Fig. 4 depicts the oscillation amplitude of bunch centroid with respect to bunch index in the $5000^{th}$ turn. The fill pattern is the same as used in Fig.3. It shows that the oscillation amplitude of the bunch tail is larger than that of the bunch head, which coincides with the experimental test results performed at SSRF storage ring [7]. Fig.5 shows beam maximum oscillation amplitude with respect to the number of turns for the bunch current of 0.3 mA at CO pressure of 0.5 nTorr, in which 600 bunches are filled in the ring. The vertical amplitude of beam is below the beam size in this case. It indicates clearly that the lower gas pressure in the beam pipe, the less growth rate of the FII.

In addition, we also simulate the growth of fast ion instability in the mini-train case, in which a long bunch train of 600 bunches is divided into 4 short bunch trains with each consisting of 150 bunches, followed by the gap of 30 RF buckets (LtrainGap =30) between the adjacent trains. The simulation result is shown in Fig. 6. The total number of bunches is 600 ($n_b \times n_{train}$=600), bunch current is 0.5 mA and the CO pressure is 1.0 nTorr. Compared to Fig.2 for a long bunch train case, it shows that the beam oscillation amplitude is below the beam size when mini-train is introduced. Fig.7 compares the results for another long bunch train fill (540 bunches in total) and mini-train fill case in which the total number of bunches are 540 in the ring. At the bunch current of 0.5 mA and CO pressure of 1 nTorr, the beam oscillation amplitude for mini-train case, in which 6 bunch trains with each consisting of 90 bunches, with the train gap of 30 RF buckets in between, is quite lower than the beam size. It indicates that the mini-train is very helpful to alleviate the growth of the FII.

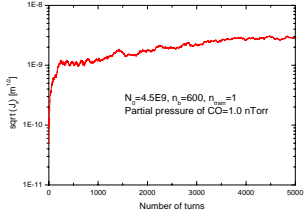

Fig.1: Beam maximum horizontal oscillation amplitude *vs*. number of turns for bunch current of 0.5 mA at CO partial pressure of 1.0 nTorr.

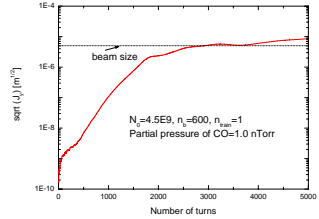

Fig.2: Beam maximum vertical oscillation amplitude *vs*. number of turns for bunch current of 0.5 mA at CO partial pressure of 1.0 nTorr.

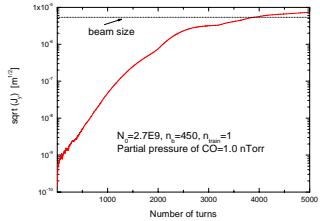

Fig.3: Beam maximum vertical oscillation amplitude *vs*. number of turns for bunch current of 0.3 mA at CO partial pressure of 1.0 nTorr.

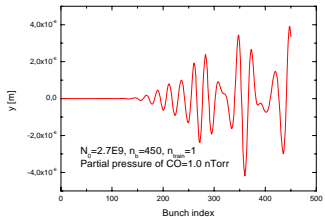

Fig. 4: The oscillation amplitude of bunch centroid *vs*. the bunch index.

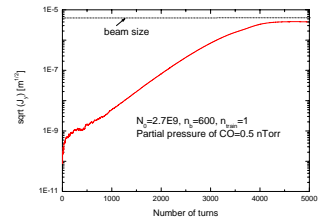

Fig. 5: Beam maximum amplitude *vs*. the number of turns for the bunch current of 0.3 mA at CO pressure of 0.5 nTorr.

Besides, we also investigate the growth of the FII due to hydrogen gas in the vacuum pipe. Fig. 8 gives the evolution of the beam maximum oscillation amplitude with respect to number of turns for bunch current of 0.5 mA at $H_2$ partial pressure of 1.0 nTorr. It shows clearly that the oscillation amplitude due to $H_2$ ions is well below the beam size. This is due to the collision ionization cross section of $H_2$ ions is much less than that of the CO ions. Except that, the small mass of hydrogen means that it can easily escape from the beam potential (critical mass of ions trapped in the beam is inversely proportional to the mass of ions).

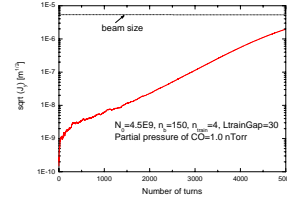

Fig.6: Beam maximum oscillation amplitude *vs*. number of turns for bunch current of 0.5 mA at CO partial pressure of 1.0 nTorr. Minitrain is introduced in the fill.

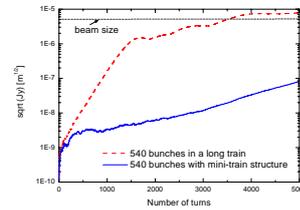

Fig.7: Beam maximum oscillation amplitude *vs*. number of turns for a long bunch trains with 540 bunches and a mini-train bunch structure with 6 short bunch trains, each train consisting of 90 bunches.

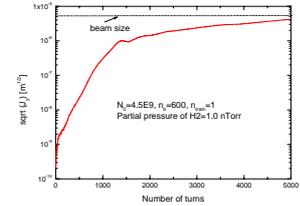

Fig.8: Beam maximum oscillation amplitude *vs*. number of turns for beam current of 0.5 mA/bunch at $H_2$ partial pressure of 1.0 nTorr.

## CONCLUSION

The fast ion instability in SSRF storage ring has been simulated by using a weak-strong code. Simulation results show that the beam vertical oscillation amplitude grows beyond the beam size for current typical fill patterns at the CO pressure of 1.0 nTorr. If a long bunch train is divided into several mini-trains with some empty RF buckets in between as clearing gaps, the growth of fast ion instability will be reduced significantly.